\documentclass[twocolumn,english,pra,aps,superscriptaddress]{revtex4-2}
\usepackage[T1]{fontenc}
\usepackage{color}
\usepackage{babel}
\usepackage{mathtools}
\usepackage{bm}
\usepackage{graphicx}
\usepackage{esint}
\usepackage{amsfonts}
\usepackage{graphicx}
\usepackage{float}
\usepackage{braket}
\usepackage{subfigure}
\usepackage{amsthm}
\usepackage{amssymb}

\usepackage[usenames,divipsnames,svgnames,table]{xcolor}
\usepackage[unicode=true,pdfusetitle,
bookmarks=true,bookmarksnumbered=false,bookmarksopen=false,
breaklinks=true,pdfborder={0 0 0},backref=false,
colorlinks=true]{hyperref}
\hypersetup{linkcolor=NavyBlue,urlcolor=NavyBlue,citecolor=NavyBlue}
\usepackage{breakurl}

\begin{document}

\title{Information Geometry and Parameter Sensitivity of Non-Hermitian Hamiltonians}

\author{Wangjun Lu}
\affiliation{Department of Maths and Physics, Hunan Institute of Engineering, Xiangtan 411104, China}
\affiliation{Institute of Engineering Education and Engineering Culture Innovation,  Hunan Institute of Engineering, Xiangtan 411104, China}
\affiliation{College of Physics and Electronic Engineering, Hengyang Normal University, Hengyang 421002, China}

\author{Zhao-Hui Peng}
\affiliation{Hunan Provincial Key Laboratory of Intelligent Sensors and Advanced Sensor Materials, and Department of Physics, Hunan University of Science and Technology, Xiangtan 411201, China}

\author{Hong Tao}
\email{hongtao@quanta.org.cn}
\affiliation{Department of Maths and Physics, Hunan Institute of Engineering, Xiangtan 411104, China}

\begin{abstract}
Information geometry is the application of differential geometry in statis-
tics, where the Fisher-Rao metric serves as the Riemannian metric on the statis-
tical manifold, providing an intrinsic property for parameter sensitivity. In this
paper, we explores the application of information geometry in the realm of non-Hermitian quantum systems, focusing on the Fisher-Rao metric as a measure of parameter sensitivity. We approximate the Lindblad master equation for non-Hermitian Hamiltonians to analyze the temporal evolution of the quantum geometric metric. Utilizing the quantum spin Ising model with an imaginary magnetic field as an exemplar, we investigate the energy spectrum and geometric metric evolution within $\mathcal{PT}$-symmetry Hamiltonians. We demonstrate that the detrimental effects of dissipation can be counteracted by introducing a control Hamiltonian, leading to improved accuracy in parameter estimation. Our work provides insights into the role of quantum control in mitigating dissipative impacts and enhancing the precision of quantum metrological tasks.

~~~

\end{abstract}

\maketitle

\section{introduction}

The manifold of coupling constants that parameterize a quantum Hamiltonian is naturally equipped with a Riemannian metric, which is pivotal for operational distinguishability as demonstrated in the works of Shannon and Zanardi~\cite{Shannon1948, Zanardi2007}. Information geometry serves as a structured methodology for probing the sensitivity of physical systems' states to parameter variations, as elucidated by Amari and others~\cite{Amari2007, Calin22014, Amari2016, Sosuke2018, Gianfrate2020, Nielsen2020, Zhang2019, Yu2019}. 
In scenarios involving non-Hermitian Hamiltonians, which are used to describe certain physical systems, phase transitions may occur, leading to sharp alterations in the system's dynamical characteristics~\cite{Zanardi2006, Cejnar2007, Zhang2020, Guo2009, Bender2013}. At these critical junctures, the system's state exhibits heightened sensitivity to fluctuations in the Hamiltonian's parameters. This sensitivity is quantifiable through the application of the Riemannian metric within the quantum state space.

From the vantage point of parameter estimation theory, as explored by Helstrom, Holevo, and others~\cite{Helstrom1976, Holevo1982, Lu2021, Liu2020, Zhang2022, Chen2022, Szczykulska2016, Humphreys2013, Gessner2018, Goldberg2021}, the parameter sensitivity of non-Hermitian Hamiltonians is a subject of considerable interest. The information geometry approach facilitates this analysis by offering a metric that quantifies the distance between Hamiltonians across varying parameter values, thereby providing a means to assess the impact of parameter changes on the system's state.

Quantum mechanics has historically concentrated on the study of closed systems, which are governed by Hermitian Hamiltonians. This focus is rooted in the foundational principles of quantum mechanics that have been in place since its inception~\cite{Barton1963}. Hermiticity of a Hamiltonian is a crucial attribute because it guarantees the conservation of probability within a quantum system and ensures that the energy spectrum is real. This is essential, as all physical measurements of a system's energy must yield real numbers~\cite{Ashida2020}.

However, in recent years, there has been a surge of interest in exploring more general Hamiltonians that are not necessarily Hermitian. This shift involves relaxing the Hermiticity condition, which opens up new avenues of research and the potential for novel physical phenomena. The exploration of non-Hermitian Hamiltonians is motivated by the desire to understand and harness the properties of systems that may exhibit non-traditional behavior, such as non-reciprocal energy transfer or the emergence of exceptional points where the usual assumptions of quantum mechanics are challenged. This broadened perspective is reshaping our understanding of quantum systems and their possible applications.

A non-Hermitian Hamiltonian is distinguished by the condition \( H \neq H^\dagger \), where \( H^\dagger \) is the adjoint of \( H \). This type of Hamiltonian is instrumental in describing dissipative processes, such as radioactive decay phenomena~\cite{Bender2007}. When employed to model a particle undergoing radioactive decay, a non-Hermitian Hamiltonian predicts a decrease in the probability of detecting the particle over time~\cite{Bender1999, Bender1998, Gao2015}. It is essential to note that a particle does not simply disappear, as this would violate the principle of probability conservation. Instead, the particle undergoes a transformation into other particles. Consequently, the non-Hermitian Hamiltonian provides a simplified, phenomenological model of decay, without delving into the specific details of the decay products.

In a groundbreaking development in 1998, it was shown that the eigenvalues of a parity-time ($\mathcal{PT}$) symmetric Hamiltonian can be entirely real, despite the lack of Hermiticity~\cite{Bender1998}. This revelation challenged the established paradigm of quantum mechanics. The $\mathcal{PT}$ symmetry allows for the existence of non-Hermitian and complex Hamiltonians, enabling a consistent framework for quantum mechanics that does not rely on Hermiticity~\cite{Bender2003, Bender2002, Miri2019, Christodoulides2018}. This insight has broadened the scope of quantum theory, highlighting the potential for alternative descriptions of physical systems that were previously thought to be constrained by the requirement of Hermiticity.

In this study, we extend the concept of quantum information geometry to non-Hermitian systems. Given that a $\mathcal{PT}$-symmetric Hamiltonian in the unbroken phase possesses real eigenvalues, it is feasible to contemplate quantum metrics within the framework of $\mathcal{PT}$-symmetric quantum systems. Our investigation commences with an analysis of the quintessential 2-dimensional $\mathcal{PT}$-symmetric Hamiltonian, illustrating the progression of its energy across both the broken and unbroken phases. Subsequently, we present the Fisher-Rao metric for the $\mathcal{PT}$ Hamiltonian, underscoring that it is indeed feasible to derive such a metric for complex quantum systems, and delve into the intricacies of parameter estimation within non-Hermitian frameworks.

For $\mathcal{PT}$-symmetric Hamiltonians, the quantum Fisher metric, as dictated by the Schrödinger equation, mirrors the behavior observed in Hermitian systems when operating under biorthogonal bases. We further derive the effective formulation of the Lindblad master equation, which encapsulates the dynamical evolution of the system, encompassing quantum jumps. From this derived Lindblad master equation, we delineate the temporal evolution of the Fisher-Rao metric. Nonetheless, it is noted that the Fisher-Rao metric is subject to fluctuations induced by dissipation.

To elucidate this phenomenon, this paper examines the quantum spin Ising model, with a particular focus on the Yang-Lee model. By incorporating an imaginary magnetic field in the x-direction, the study reveals that dissipation exerts a more pronounced impact on the Fisher-Rao metric. To counteract the repercussions of the imaginary magnetic field, we introduce a control Hamiltonian designed to ameliorate its effects, thereby enhancing the precision of parameter estimation.

This work is organized as follows. In Sec.~\ref{sec:nhh}, Introduction to the 2-dimensional $\mathcal{PT}$-Symmetric Hamiltonian. In Sec.~\ref{Sec:QFR}, Quantum Information Geometry for $\mathcal{PT}$-Symmetric Hamiltonians. In Sec.~\ref{Sec:nhhdy}, we provide a brief overview of the dynamics with the non-Hermitian Hamiltonian. In Sec.~\ref{Sec:example}, We compute the Fisher-Rao metric with the Yang-Lee model. We provide a summary in Sec.~\ref{Sec:conclution}.
 
\section{The $\mathcal{PT}$-symmetry Hamiltonian}\label{sec:nhh}
In recent years, the investigation of non-Hermitian Hamiltonians within the framework of quantum theory has garnered significant attention, primarily fueled by their ability to model systems exhibiting remarkable properties that adhere to the principle of $\mathcal{PT}$-symmetry. This paradigm shift challenges the long-held dogma in quantum mechanics, where Hermiticity, especially of Hamiltonians, has been paramount for guaranteeing real energy spectra and unitary time evolution.

Conventionally, quantum mechanics has relied heavily on the Hermiticity of operators, particularly Hamiltonians, as a cornerstone to ensure the reality of energy eigenvalues and the preservation of unitarity during time evolution. However, seminal works by Bender and collaborators have fundamentally challenged this notion, demonstrating that the stricter Hermiticity condition is not an absolute necessity for these properties to hold~\cite{Bender1998, Bender1999, Bender2007}. Instead, they have introduced the weaker condition of $\mathcal{PT}$-symmetry, which they have shown to be sufficient to ensure a real energy spectrum and unitary time evolution for a specific class of non-Hermitian Hamiltonians. This discovery has paved the way for exploring the rich physics and novel phenomena that arise from these unconventional Hamiltonians, broadening the horizons of quantum theory.

The Hamiltonian $H$ is designated as $\mathcal{PT}$-symmetric if it remains invariant under the joint operation of the parity operator $\mathcal{P}$ and the time reversal operator $\mathcal{T}$. This symmetry is mathematically formulated as:
$ H = H^{\mathcal{PT}}$, 
where $H^{\mathcal{PT}}$ denotes the $\mathcal{PT}$-transformed Hamiltonian.
The parity operator $\mathcal{P}$ executes spatial reflections, which in the context of one-dimensional systems, translates to $x \rightarrow -x$. It possesses unitary properties, ensuring the preservation of the norm of any state upon its application. Consequently, the square of the parity operator is the identity operator, signified as $\mathcal{P}^2 = \openone$.

The time reversal operator $\mathcal{T}$, on the other hand, reverses the flow of time and induces a complex conjugation on any complex-valued quantities. Specifically, it maps $i \rightarrow -i$. This operator is anti-unitary, meaning that it reverses the phase of any state it acts upon while maintaining its magnitude. Similarly, the square of the time reversal operator is also the identity operator, expressed as $\mathcal{T}^2 = \openone$.

The conjunction of these two operators, $\mathcal{PT}$, forms the basis for identifying a subset of non-Hermitian Hamiltonians that exhibit remarkable properties akin to those of conventional Hermitian Hamiltonians, such as real energy spectra and unitary time evolution, under certain conditions.

The parity and time reversal operators exhibit a fundamental property of commutativity, expressed as $\mathcal{P}\mathcal{T} = \mathcal{T}\mathcal{P}$. This commutativity ensures that the order in which these operators are applied is immaterial, simplifying the analysis of systems exhibiting $\mathcal{PT}$ symmetry.

For the position operator $\hat{x}$, the parity operator $\mathcal{P}$ introduces a sign flip, reflecting the coordinate system: $ \mathcal{P}\hat{x}\mathcal{P} = -\hat{x} $. Conversely, the time reversal operator $\mathcal{T}$ does not directly alter the position operator for real-valued positions, as it neither changes the sign of $\hat{x}$ nor introduces complex conjugations in a non-trivial manner. Consequently, under the combined $\mathcal{PT}$ transformation, the position operator remains transformed as:
$$ \mathcal{PT}\hat{x}(\mathcal{PT})^{-1} = -\hat{x}. $$

In the case of the momentum operator $\hat{p}$, the time reversal operator $\mathcal{T}$ reverses the direction of motion, leading to a sign change in the momentum: $ \mathcal{T}\hat{p}\mathcal{T}^{-1} = -\hat{p} $. Conversely, the parity operator $\mathcal{P}$ does not directly modify the momentum operator, neither by changing its sign nor introducing complex conjugations. Thus, under the joint $\mathcal{PT}$ operation, the momentum operator similarly transforms as:
$$ \mathcal{PT}\hat{p}(\mathcal{PT})^{-1} = -\hat{p}. $$

When a Hamiltonian $H$ possesses $\mathcal{PT}$-symmetry, it signifies that the Hamiltonian remains invariant under the combined action of the parity operator $\mathcal{P}$ and the time reversal operator $\mathcal{T}$. This invariance has profound implications in quantum mechanics, particularly in the realm of non-Hermitian Hamiltonians, which, despite their non-standard properties, can exhibit remarkable features such as real energy spectra due to their adherence to $\mathcal{PT}$-symmetry.

In this section, we present the theoretical framework for analyzing a specific two-level quantum system governed by a non-Hermitian Hamiltonian, building upon the pioneering works of Bender and his collaborators \cite{Bender1998, Bender2007}. The Hamiltonian of this system, given by
\begin{equation}\label{eq:nHH_model}
  H = s\sigma^x - ir\sigma^z,
\end{equation}
where $s$ and $r$ are real parameters, and $\sigma^x$ and $\sigma^z$ are the standard Pauli matrices, serves as an exemplary model to illustrate the phenomenon of $\mathcal{PT}$-symmetry breaking. Here, the parity operator $\mathcal{P}$ is defined as $\sigma^x$, which flips the sign of the $x$-component of the spin, while the time-reversal operator $\mathcal{T}$ involves complex conjugation.

To delve into the behavior of this system, we calculate its eigenvalues, which are expressed as
\begin{equation}
   E_{\pm} = \pm\sqrt{s^2 - r^2}.
\end{equation}
The nature of the square root in this expression leads to the identification of two distinct parametric regions:

Broken $\mathcal{PT}$-symmetry regime ($s^2 < r^2$): In this region, the eigenvalues $E_{\pm}$ form a complex-conjugate pair, indicating the spontaneous breaking of $\mathcal{PT}$-symmetry. This feature is a hallmark of non-Hermitian systems, where the reality of the energy spectrum is no longer guaranteed. 

Unbroken $\mathcal{PT}$-symmetry regime ($s^2 > r^2$): In this regime, the eigenvalues remain real, signifying the preservation of $\mathcal{PT}$-symmetry. Within this parameter space, the system behaves similarly to conventional Hermitian systems, with real energy levels and corresponding eigenstates.

To further our analysis within the unbroken $\mathcal{PT}$-symmetry region, we explicitly construct the simultaneous eigenstates of the operators $H$ and $\mathcal{PT}$. These states are given by
\begin{equation}
  \left|E_{\pm}\right\rangle = n_{\pm}\left(\begin{array}{c} 1 \\ i r \pm \sqrt{s^2 - r^2} \end{array}\right),
\end{equation}
where the normalization constants $n_{\pm}$ are chosen to ensure proper normalization of the states, satisfying $n_+n_+^* = \frac{1}{2\sqrt{s^2 - r^2}}$. These eigenstates provide insight into the system's dynamics and its response to perturbations that might disrupt the $\mathcal{PT}$-symmetry.

\begin{figure}[tp]
  \includegraphics[width=\columnwidth]{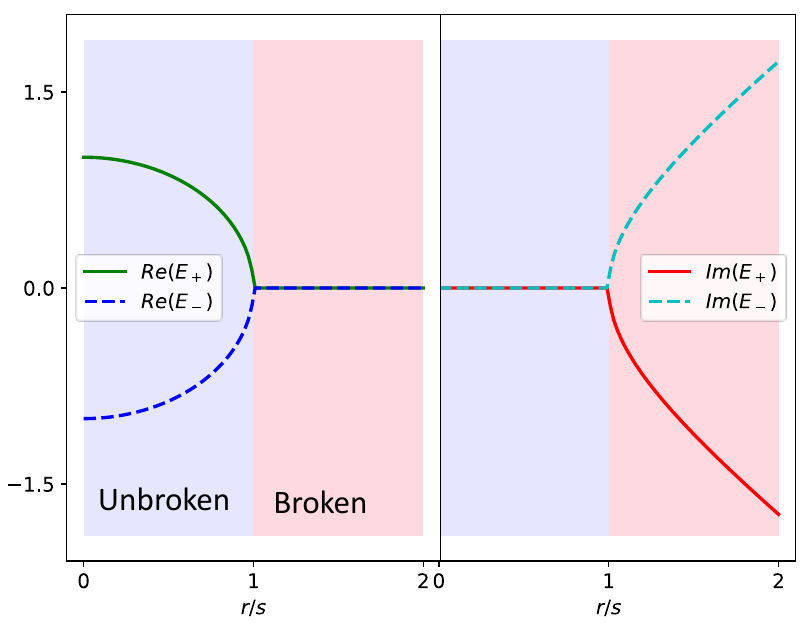} \centering
  \caption{Spectral properties of the non-Hermitian Hamiltonian. 
  Imaginary and real parts of the eigenvalues in Eq.~(\ref{eq:nHH_model})
  as a function of $r/s$. } \label{Fig:_value}
  \end{figure}

To visually illustrate the behavior of the energy eigenvalues, we propose plotting the imaginary and real parts of $E_{\pm}$ as functions of the parameters $s$ and $r$, as depicted in Fig.~\ref{Fig:_value}. These plots will reveal a clear transition from the unbroken to the broken $\mathcal{PT}$-symmetry region as $s^2$ crosses the threshold $r^2$.

At the critical juncture characterized by the condition $r/s = 1$, the eigenvalues converge, resulting in a degenerate state with $E_+ = E_- = 0$. This exceptional point delineates the boundary between the unbroken and broken $\mathcal{PT}$-symmetry phases.

In the parameter domain defined by $s^2 > r^2$, the eigenvalues $E_{\pm}$ manifest as purely real, thereby indicating the unbroken $\mathcal{PT}$-symmetry. This observation is corroborated through an examination of the $\mathcal{PT}$-inner product of the corresponding eigenstates, which adhere to the orthogonality relations:
\begin{align*}
\left\langle E_{\pm}|\mathcal{P} \mathcal{T}| E_{\pm}\right\rangle = \pm 1, ~~
\left\langle E_{\pm}|\mathcal{P} \mathcal{T}| E_{\mp}\right\rangle = 0,
\end{align*}
These relations substantiate that the eigenstates $|E_{\pm}\rangle$ are orthogonal with respect to the $\mathcal{PT}$-inner product, thereby confirming their status as bona fide eigenstates (with eigenvalue $+1$) or pseudo-eigenstates (with eigenvalue $-1$) of the $\mathcal{PT}$ operator. The orthogonality, in conjunction with the real nature of the eigenvalues, affirms the preservation of $\mathcal{PT}$-symmetry within this parameter space.

In contradistinction, when the condition $s^2 < r^2$ is satisfied, the eigenvalues $E_{\pm}$ emerge as a complex-conjugate pair, signifying the spontaneous breaking of $\mathcal{PT}$-symmetry. Within this broken phase, the $\mathcal{PT}$-inner product no longer guarantees the orthogonality of the eigenstates, imparting the system with distinctive non-Hermitian attributes that set it apart from Hermitian quantum systems

Indeed, by comprehensively examining both the eigenvalues and the $\mathcal{PT}$-inner products of the eigenstates, we can definitively ascertain whether the $\mathcal{PT}$-symmetry of a given Hamiltonian is preserved or broken. This approach allows for a direct comparison between the imaginary parts, Im($E_i$), and the real parts, Re($E_i$), of the eigenvalues, offering a comprehensive understanding of their behavior. The analysis of the eigenvalues $E_{\pm}$ and the associated $\mathcal{PT}$-inner products serves as a vital tool in elucidating the underlying $\mathcal{PT}$-symmetry properties of the system.

\section{ Information  Geometry with complex systems}\label{Sec:QFR}

In the quest to unravel the geometrical intricacies of parametric subspaces within the quantum state space, the Fubini-Study metric emerges as a pivotal analytical instrument. This metric is uniquely suited to delineate the intrinsic properties of the quantum state space, which is conventionally construed as the ensemble of rays projecting from the origin within the Hilbert space, a construct more formally known as the complex projective space \cite{Brody2013}.

The quantum state space, mathematically represented as the complex projective space \( \mathbb{C}P^n \), encapsulates the set of all one-dimensional subspaces of an \( n \)-dimensional Hilbert space. This formulation is pivotal as it disregards the global phase of quantum states, thus focusing on the physical content that is invariant under phase transformations.

The Fubini-Study metric is also known as the Fisher-Rao metric or the quantum Fisher information, denoted by \( g_{FS} \), quantifies the proximity between quantum states in this space. It is derived from the Riemannian metric adapted to the complex projective space and is invariant under the unitary transformations of the Hilbert space. This invariance is of paramount importance, as it ensures that the geometrical properties of the quantum state space are preserved under the action of quantum dynamics described by unitary operators.

Moreover, the Fisher-Rao metric is instrumental in the field of quantum information theory, where it plays a crucial role in quantum estimation theory. It underpins the formulation of quantum limits to parameter estimation, offering insights into the precision with which parameters can be estimated in quantum systems.

\subsection{  The Hermitian system}
In the realm of quantum information geometry, the assessment of the infinitesimal distance \( \mathrm{d}s \) between two quantum states that are infinitesimally close, specifically \( |\psi\rangle \) and \( |\psi + \mathrm{d}\psi\rangle \), is a fundamental task. The latter state can be expressed as \( |\psi + \mathrm{d}\psi\rangle = |\psi\rangle + |\mathrm{d}\psi\rangle \), where \( |\mathrm{d}\psi\rangle \) is the infinitesimal variation of the state.

To derive the Riemannian metric that quantifies this distance, we retain terms that are second-order in the differential \( \mathrm{d}\psi \), thereby capturing the essence of the geometric structure of the quantum state space. This approach is pivotal for investigating the geometric properties of quantum states as they undergo parametric transformations.

Let us consider a quantum state \( |\psi\rangle = |\psi(\theta^i)\rangle \) that depends smoothly on a set of parameters \( \{\theta^i\} \), and is normalized such that \( \langle \psi | \psi \rangle = 1 \) for all values of \( \{\theta^i\} \). The infinitesimal variation of the state \( |\mathrm{d}\psi\rangle \) can be expressed in terms of the partial derivatives with respect to these parameters, employing the Einstein summation convention: \( |\mathrm{d}\psi\rangle = \partial_i \psi \, \mathrm{d}\theta^i \), where \( \partial_i \equiv \frac{\partial}{\partial \theta^i} \) and the summation over repeated indices is implied.

The quantum Fisher-Rao metric \( g_{ij} \) on the Riemannian manifold, induced by the ambient Fubini-Study metric, is then derived as follows:
\begin{align}\label{EQ:ds}
  \mathrm{d}s^2 &= g_{ij} \, \mathrm{d}\theta^i \mathrm{d}\theta^j \\ \nonumber
  &= \left( \langle \partial_i \psi | \partial_j \psi \rangle - \langle \psi | \partial_i \psi \rangle \langle \partial_j \psi | \psi \rangle \right) \mathrm{d}\theta^i \mathrm{d}\theta^j.  
\end{align}

This metric tensor \( g_{ij} \) encapsulates the local geometry of the manifold and is explicitly given by:
\begin{equation}
g_{ij} = \langle \partial_i \psi | \partial_j \psi \rangle - \langle \psi | \partial_i \psi \rangle \langle \partial_j \psi | \psi \rangle.
\end{equation}

This formulation of the quantum Fisher-Rao metric serves as a cornerstone for analyzing the curvature and the geometrical properties of the manifold of quantum states. It provides a means to quantify the distinguishability of quantum states and is instrumental in quantum estimation theory, where it plays a crucial role in determining the precision limits of parameter estimation in quantum systems.

\subsection{  The non-Hermitian system}

In this work, we extend the geometric analysis of parametric spaces from the realm of real Hilbert spaces to the complex domains that underpin the state space of quantum mechanics. This extension necessitates a nuanced approach, as it involves modifications beyond a mere transposition of concepts from real to complex spaces.

In the context of Hermitian Hamiltonians, the quantum information geometry naturally manifests as a Riemannian manifold. This is predicated on the fact that Hermitian operators possess real eigenvalues and can be associated with a complete set of orthogonal eigenstates. This orthogonality allows for the construction of a well-defined metric structure on the space of quantum states, facilitating a geometric interpretation of quantum state transformations.

The scenario becomes more complex with $\mathcal{PT}$-symmetric Hamiltonians. These are non-Hermitian operators that, in certain regions of the parameter space known as the unbroken $\mathcal{PT}$-symmetric region, can exhibit a real eigenvalue spectrum. In this region, the eigenstates of the Hamiltonian form a biorthogonal basis. Specifically, the left ($\langle \tilde{\psi}(t)|$) and right ($|\psi(t)\rangle$) eigenstates satisfy the biorthogonality condition:
\begin{align}
\langle \tilde{\psi}_n(t) | \psi_m(t) \rangle = \delta_{nm},
\end{align}
where $\delta_{nm}$ is the Kronecker delta.

For a non-Hermitian Hamiltonian $H$, the eigenvalue equation provides a framework to describe the system's behavior:
\begin{align}
H |\psi\rangle = E |\psi\rangle, \quad H^\dagger |\tilde{\psi}\rangle = \tilde{E} |\tilde{\psi}\rangle,
\end{align}
where $E$ and $\tilde{E}$ are the eigenvalues, and $|\tilde{\psi}\rangle$ is related to $|\psi\rangle$ through the non-Hermitian Hamiltonian $H$. Assuming the system evolves according to the non-Hermitian Schrödinger equation:
\begin{align}\label{eq:schr}
  H|\psi(t)\rangle = i \partial_t|\psi(t)\rangle, ~~  H^{\dagger}|\tilde{\psi}(t)\rangle = i \partial_t|\tilde{\psi}(t)\rangle.
\end{align}
and given the initial normalization $\langle\tilde{\psi}(0) \mid \psi(0)\rangle = 1$, we derive the orthogonality condition for quantum states in a non-Hermitian quantum system:
\begin{align}
\langle\tilde{\psi}(t) \mid \psi(t)\rangle = 1.
\end{align}

We consider a pure state $|\psi(\theta)\rangle$ parameterized by a variable $\theta$, with $\langle\psi(\theta)|$ denoting its Hermitian conjugate. In the complex Hilbert space, the condition $\partial_\theta \langle \psi(\theta)|\psi(\theta)\rangle = 0$ does not necessarily imply that $\langle \partial_\theta \psi(\theta)|\psi(\theta)\rangle = 0$, due to the potential non-trivial phase factor associated with the complex state. This subtlety is crucial when analyzing the geometric properties of quantum states in complex spaces.

For an $N$-dimensional $\mathcal{PT}$-symmetric quantum system, the state $|\psi(t)\rangle$ (and its dual $|\tilde{\psi}(t)\rangle$) is parameterized by an estimation parameter $\theta$. The Fisher-Rao metric, an alternative expression for the Fubini-Study metric, can be used to investigate the metric geometry of the eigenstates of non-Hermitian Hamiltonians:
\begin{align}
g_{ij} = \langle \partial_i \tilde{\psi}(t) | \partial_j \psi(t) \rangle - \langle \partial_i \tilde{\psi}(t) | \psi(t) \rangle \langle \tilde{\psi}(t) | \partial_j \psi(t) \rangle.
\end{align}
Here, $\partial_i$ and $\partial_j$ represent partial derivatives with respect to the parameters $i$ and $j$ that parameterize the quantum states $|\psi(t)\rangle$ and $|\tilde{\psi}(t)\rangle$, respectively. This metric captures the geometric structure of the parameter space in which the eigenstates evolve under non-Hermitian dynamics.

Despite the non-Hermitian nature of the Hamiltonian, the Fisher-Rao metric provides a meaningful measure of the distance between neighboring quantum states in the parameter space, especially in the unbroken $\mathcal{PT}$-symmetric region where the eigenvalue spectrum is real. This allows for a geometric analysis of quantum information even in the presence of non-Hermitian dynamics. In the subsequent sections, we will provide an illustrative example to demonstrate the variations of the Fisher-Rao metric as the system parameters are varied within the $\mathcal{PT}$-symmetry phase.

\subsection{Example : two-level non-Hermitian systems}
\begin{figure}[tp]
  \includegraphics[width=\columnwidth]{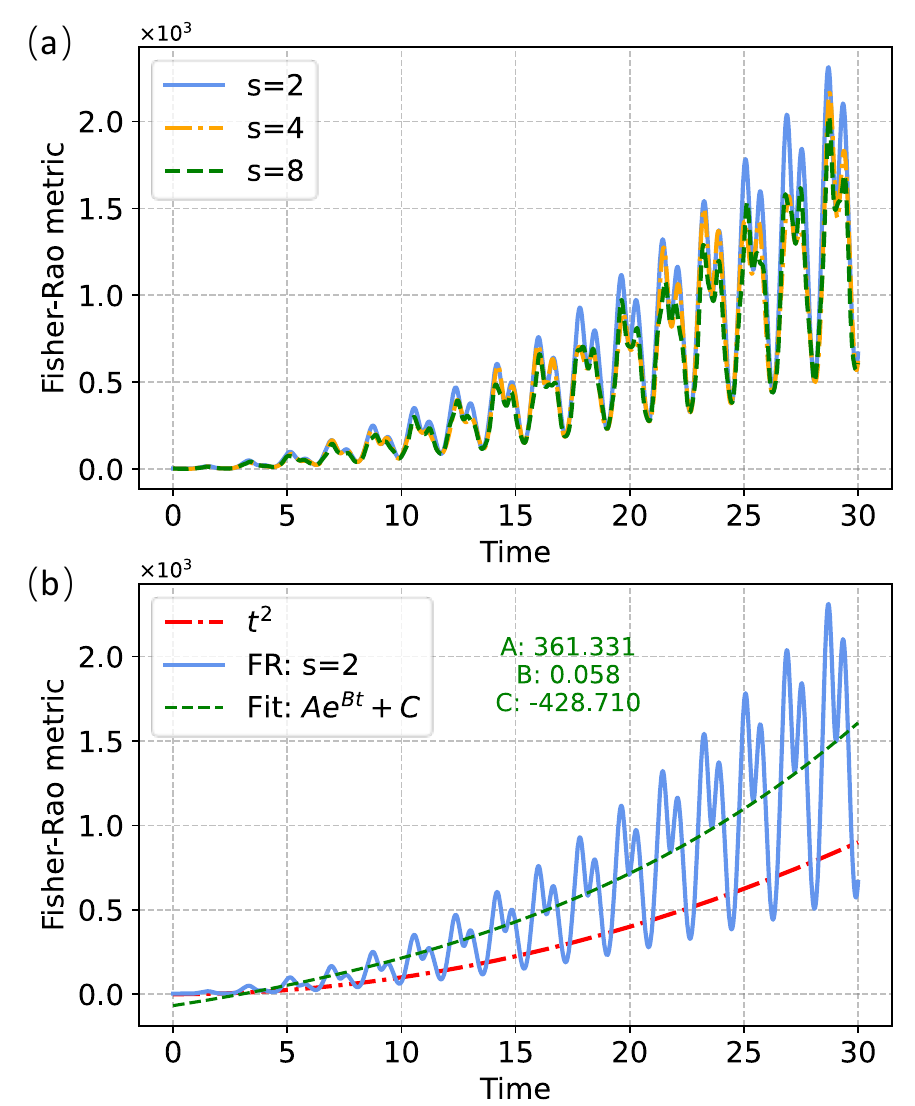} \centering
  \caption{Evolution of the Fisher-Rao Metric (FR) Governed by the Hamiltonian Eq.~(\ref{eq:nHH_model}): (a) In the unbroken phase, the parameter $s$ exceeds $r$. The distinct colored lines portray the evolution for various values of $s$, namely $s=2$, $s=4$, and $s=6$. (b) Analysis of the Fisher-Rao Metric Evolution: The blue curve delineates the temporal progression of the Fisher-Rao metric. The red dotted line represents the square of time, serving as a benchmark. The green line presents a fitting curve that encapsulates the trend of the Fisher-Rao metric's evolution, with A, B, C indicating specific fitted values.} \label{fig:FR}
  \end{figure}
In this study, we consider a non-Hermitian Hamiltonian for the description of a quantum system governed by the dynamics outlined in Eq.~(\ref{eq:nHH_model}). We commence by initializing the system in a specific single-qubit state, denoted as
$$ |\psi_0\rangle = \frac{1}{\sqrt{2}}(|0\rangle + |1\rangle), $$
which represents an equal superposition of the computational basis states $|0\rangle$ and $|1\rangle$. Subsequently, we investigate the non-unitary time evolution of this quantum system, which is governed by the Schrödinger equation [Eq.~(\ref{eq:schr})]. 

The explicit expressions for the time-evolved quantum states $|\psi(t)\rangle$ and $|\tilde{\psi}(t)\rangle$ are given as follows:
\begin{align*}
|\psi(t)\rangle &= \frac{1}{\sqrt{2}\mathcal{N}} \left(\begin{array}{c}
E_+\cos (t E_+) - \left(is -r\right) \sin (t E_+) \\
E_+ \cos (t E_+) - \left(is +r\right)\sin (t E_+)
\end{array}\right), \\
|\tilde{\psi}(t)\rangle &= \frac{1}{\sqrt{2}\mathcal{N}} \left(\begin{array}{c}
E_+\cos (t E_+) + \left(is +r\right) \sin (t E_+) \\
E_+ \cos (t E_+) + \left(is -r\right) \sin (t E_+)
\end{array}\right),
\end{align*}
where the normalization factor $\mathcal{N}$ ensures that the quantum states $|\Psi(t)\rangle$ and $|\tilde{\Psi}(t)\rangle$ remain properly normalized at all times. Specifically, $\mathcal{N}$ is defined as
$$ \mathcal{N} = \sqrt{s^2 - r^2 \cos (2 \sqrt{s^2-r^2} t)}, $$
which accounts for the non-unitary nature of the dynamics characterized by the parameters $r$ and $s$.

Employing the QuanEstimation package~\cite{Zhang2022}, we compute the quantum Fisher-Rao metric (FRM) in conjunction with the specified Hamiltonian. As showcased in Fig.~\ref{fig:FR}, our analysis unveils that within the unbroken regime, the Fisher-Rao metric experiences an exponential ascent with the elapse of time. Notably, Fig. (a) illustrates that variations in the parameter $s$ result in only minor modifications to the Fisher-Rao metric's overall trend, implying a restricted influence of $s$ on the metric's evolution.

Furthermore, Fig. (b) presents a fitting curve for the Fisher-Rao metric, complemented by a comparative assessment of its precision boundaries. Strikingly, for systems under the purview of $\mathcal{PT}$-symmetric Hamiltonians, a significant augmentation in the precision of parameter estimation is discerned. This revelation accentuates the efficacy of the Fisher-Rao metric in quantifying the sensitivity of system parameters, thereby facilitating enhanced accuracy in the estimation of parameters within $\mathcal{PT}$-symmetric systems.

\section{The dynamics of the non-Hermitian system}\label{Sec:nhhdy}

We explore the distinctive properties of non-Hermitian Hamiltonians that set them apart from their conventional Hermitian counterparts in quantum systems. A salient feature that has garnered significant research attention is the presence of real energy eigenvalues in the unbroken $\mathcal{PT}$-symmetry region, even when the Hamiltonian itself is non-Hermitian. We now turn our focus to the dynamical processes inherent in non-Hermitian systems.

Any non-Hermitian Hamiltonian \( H \) can be decomposed into its Hermitian (\( H_{+} \)) and anti-Hermitian (\( H_{-} \)) components, as per the formulation in \cite{Sergi2011}:
\begin{align}
H_{+} = \frac{1}{2}(H + H^{\dagger}), \quad H_{-} = \frac{1}{2}(H - H^{\dagger}).
\end{align}

Given \( |\Psi(t)\rangle \) as the state evolved under the non-Hermitian Hamiltonian, the non-unitary temporal evolution of the non-normalized density matrix \( \rho(t) \) is governed by the equation presented in \cite{Sergi2015,Lindblad1976}:
\begin{equation}
\frac{d}{d t} \rho(t) = -\frac{i}{\hbar} [H_{+}, \rho(t)] + \frac{i}{\hbar} \{H_{-}, \rho(t)\},
\end{equation}
where \( [\cdot, \cdot] \) and \( \{\cdot, \cdot\} \) denote the commutator and anti-commutator, respectively. For the sake of simplicity, we set the Planck constant \( \hbar \) to unity throughout this analysis.

We incorporate the Liouvillian framework, which accounts for the dynamics of open quantum systems without quantum jumps, drawing from the seminal works of Minganti et al. \cite{Minganti2019}, Daley et al. \cite{Daley2014}, Brody et al. \cite{Brody2012}, and the pioneering experiments by Nagourney et al. \cite{Nagourney1986} and Sauter et al. \cite{Sauter1986}.

To fully describe the quantum dynamics of a dissipative system within the Lindblad formalism, we introduce the quantum jump term \( \mathcal{J}(\Gamma) \) into the dynamics. We then derive the evolution of the density matrix \( \rho(t) \) for a two-dimensional non-Hermitian Hamiltonian as:
\begin{equation}\label{eq:discrete_lindblad}
\rho(t+\tau) = \rho(t) - i\tau [H_{+}, \rho(t)] + \tau \mathcal{D}[\Gamma] \rho(t),
\end{equation}
where \( \mathcal{D}[\Gamma] = \Gamma \cdot \Gamma^{\dagger} - \frac{1}{2} \left\{ \Gamma^{\dagger} \Gamma, \cdot \right\} \) represents the Lindbladian dissipator. The quantum jump superoperator is given by \( \mathcal{J}(\Gamma) = \Gamma \rho(t) \Gamma^\dagger \), although it is not explicitly used in the discrete evolution equation.

We reformulate the discrete evolution equation into its continuous differential form, yielding the Lindblad master equation:
\begin{equation}\label{eq:master_equation}
\frac{\partial \rho(t)}{\partial t} = -i[H_+, \rho(t)] + \mathcal{D}[\Gamma] \rho(t).
\end{equation}
This master equation encapsulates the comprehensive description of the dissipative quantum dynamics under the influence of the non-Hermitian Hamiltonian.

\subsection{Fisher-Rao metric with quantum Lindblad dynamics}

\begin{figure}[tp]
  \includegraphics[width=\columnwidth]{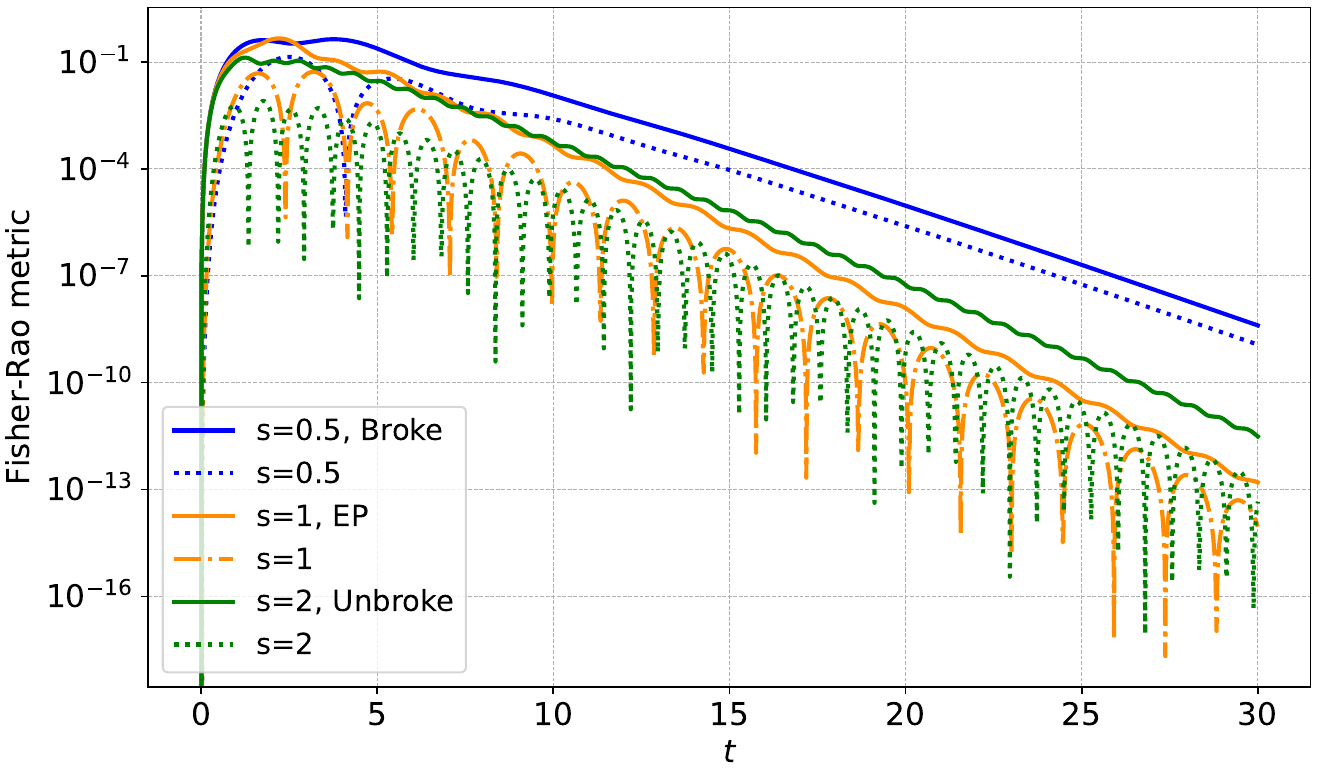} \centering
  \caption{ The evolution of the Fisher-Rao metric, under the influence of the non-Hermitian Hamiltonian specified in Equation (\ref{eq:nHH_model}), is scrutinized within the QuanEstimation framework \cite{Zhang2022}, incorporating a parameter $s$. This study focuses on the evolution patterns for three distinct ratios of $r$ to $s$: characterized by a blue line depicting $r/s = 0.5$ in the symmetry-broken regime, an orange line indicating $r/s = 1$ at the exceptional point, and a green line showcasing $r/s = 2$ in the unbroken symmetry region.  }\label{Fig_qfi_linbdlad}
  \end{figure}
Subsequently, we calculate the quantum Fisher-Rao metric, leveraging the dynamics dictated by the non-Hermitian Hamiltonian in Equation~(\ref{eq:discrete_lindblad}), using the QuanEstimation package \cite{Zhang2022}. The evolution of the Fisher-Rao metric with respect to the estimated parameter \( s \) is depicted in Fig.~\ref{Fig_qfi_linbdlad}. The horizontal axis represents time, while the vertical axis displays the logarithm of the Fisher-Rao metric. Solid lines of varying colors correspond to the quantum Fisher information for different values of \( s \). Dashed lines represent the Fisher-Rao metric under measurement conditions for the same values of \( s \), indicative of the classical Fisher information.

From Fig.~\ref{Fig_qfi_linbdlad}, it is evident that the Fisher-Rao metric exhibits an initial increase followed by a decrease over time, attributable to the dissipative effects induced by the non-Hermitian terms during the quantum state evolution. Furthermore, the Fisher-Rao metric in the symmetry-broken regime is observed to exceed that in the unbroken symmetry region, highlighting the impact of $\mathcal{PT}$-symmetry on the metric's behavior.

In the domain of quantum metrology, strategies to mitigate the adverse effects of dissipation on the Fisher-Rao metric commonly involve enhancing the control Hamiltonian, thereby improving the precision of parameter estimation in non-Hermitian quantum systems.

\subsection{Precision enhanced with quantum control}

In the domain of quantum metrology, quantum control techniques have emerged as a formidable arsenal for enhancing the precision of measurements and countering the adverse effects of dissipation inherent in parameter estimation processes. The inherent controllability within quantum metrological setups allows for the strategic implementation of such advanced strategies, thereby optimizing the performance of quantum sensors.

As depicted in the analysis presented in Fig.~\ref{Fig_qfi_linbdlad}, the Fisher-Rao metric exhibits a consistent trend across varying values of the parameter \( s \). Within the framework of the dynamics governed by the Lindblad equation (Eq.~(\ref{eq:discrete_lindblad})), which incorporates a non-Hermitian Hamiltonian (Eq.~(\ref{eq:nHH_model})), the integration of a control Hamiltonian \( H_c \) significantly bolsters the estimation process. The control Hamiltonian is articulated as:
\begin{equation}
  H_c = \sum_{k \in \{x, y, z\}} u_k S_k,
\end{equation}
where \( u_k \) are the control coefficients and \( S_k = \sum_j^N \sigma_j^k \) with \( N \) being the total number of spins. For notational simplicity, \( H_c \) can be expressed in the compact form:
\begin{equation}
  H_c = u_x \sigma^x + u_y \sigma^y + u_z \sigma^z.
\end{equation}

\begin{figure}[tp]
  \centering
  \includegraphics[width=\columnwidth]{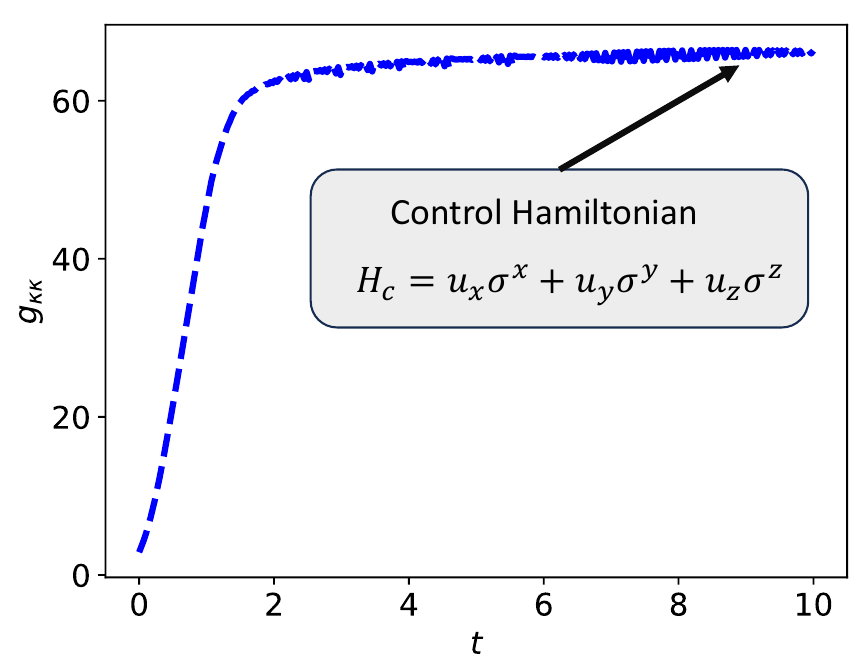}
  \caption{The performance of the control-enhanced Fisher-Rao metric is analyzed within the dynamics governed by the Lindblad master equation, as presented in Eq.~(\ref{eq:discrete_lindblad}). The true values of the parameters are set to \( r=1 \), \( s=0.2 \).} \label{Fig_ctrl-qfi}
\end{figure}

Initiating the simulation at time \( T=10 \) with the initial guess for the control parameters \( \{u_x, u_y, u_z\} \) set to zero, Figure~\ref{Fig_ctrl-qfi} reveals that the Fisher-Rao metric converges to its maximum value. This convergence underscores the effectiveness of the control method in enhancing the precision of parameter estimation, even within the context of systems governed by non-Hermitian Hamiltonians.

The essence of this approach lies in the optimization of the control parameters \( u_k \) to direct the system's evolution in a manner that maximizes the information content encoded in the quantum Fisher information (QFI). Given that the QFI is directly linked to the Fisher-Rao metric and, by extension, the precision of parameter estimation, this strategy leverages the controllability of quantum systems to counteract dissipation and elevate the overall performance of quantum metrological protocols.

The integration of quantum control within the framework of quantum metrology not only mitigates the detrimental effects of dissipation but also significantly enhances the precision of parameter estimation. This approach exemplifies the potential of quantum technologies in advancing the frontiers of measurement science.

\section{Numerical Simulation }\label{Sec:example}

We delve into the quantum Ising spin chain model, enriched with both a magnetic field in the z-direction and an additional longitudinal imaginary field along the x-axis. This model, initially introduced in \cite{Castro2009}, serves as a discrete lattice realization of the renowned Yang-Lee model \cite{Gehlen1999}, celebrated for its non-Hermitian nature and its wealth of physical properties. The Hamiltonian governing the dynamics of this system is defined as:
\begin{equation}\label{eq:YLmodel}
  H(\lambda, \kappa) = -\frac{1}{2} \sum_{j=1}^N \left( \sigma_j^z + \lambda \sigma_j^x \sigma_{j+1}^x + i \kappa \sigma_j^x \right),
\end{equation}
where \( \lambda, \kappa \in \mathbb{C} \) are the complex parameters, and \( N \) is the number of spins in the chain. The Hamiltonian operates within the Hilbert space \( \left(\mathbb{C}^2\right)^{\otimes N} \), which is the configuration space of the spin chain. The Pauli matrices \( \sigma_i^{x, y, z} \) act on the \( i \)-th site of the chain, with the identity matrix \( \openone \) filling the remaining positions in the tensor product.

The non-Hermitian character of this model provides a unique platform for exploring exotic quantum phenomena, such as exceptional points and parity-time symmetry breaking. By varying the parameters \( \lambda \) and \( \kappa \), we can tune the system's properties and potentially reveal intriguing dynamical behaviors and phase transitions not accessible in conventional Hermitian systems.

To facilitate our analysis, we decompose the Hamiltonian into its Hermitian and anti-Hermitian components:
\begin{equation}
  H(\lambda,\kappa) = H_0(\lambda) - i H_1(\kappa),
\end{equation}
where \( H_0(\lambda) \) and \( H_1 \) are Hermitian operators, and \( \lambda \) and \( \kappa \) are real constants. These components are expressed as:
\begin{align}
  H_0(\lambda) &= -\frac{1}{2} \sum_{j=1}^N \left( \sigma_j^z + \lambda \sigma_j^x \sigma_{j+1}^x \right), \\
  H_1 (\kappa) &= \frac{1}{2} \sum_{j=1}^N  \kappa \sigma_j^x.
\end{align}

The quantum dynamics of dissipative systems are comprehensively described within the Lindblad formalism. The standard Lindbladian dissipator for this model is given by:
\begin{equation}
  \Gamma = \sqrt{2H_1}.
\end{equation}

\begin{figure}[tp]
  \centering
  \includegraphics[width=\columnwidth]{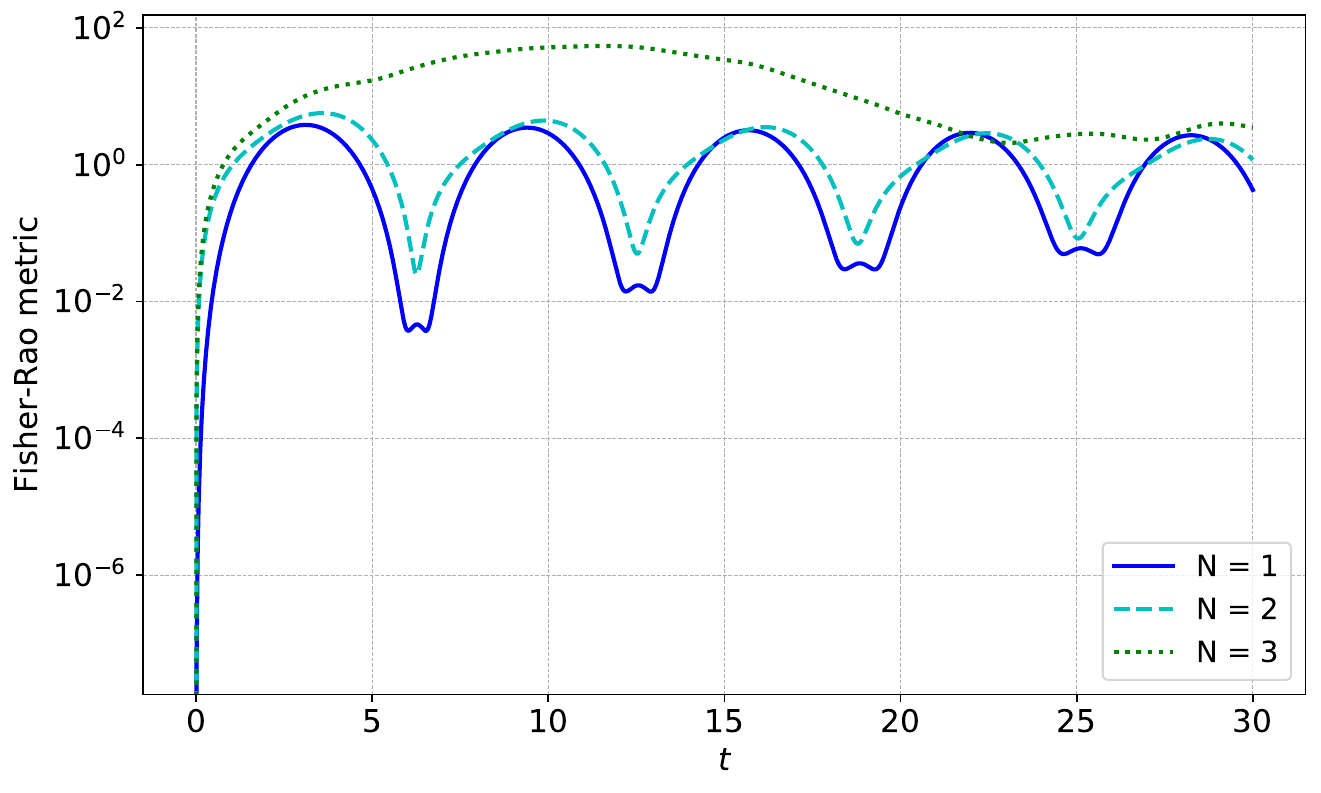}
  \caption{The non-Hermitian quantum spin model exhibits temporal evolution of the Fisher-Rao metric. The color-coded lines in the graphical representation depict the distinct evolutionary trajectories of the Fisher-Rao metric, corresponding to the values of \( N = 1, 2, \) and \( 3 \), respectively.} \label{Fig_qfi_ising}
\end{figure}

As illustrated in Fig.~\ref{Fig_qfi_ising}, we observe that the Fisher-Rao metric gradually decreases with dynamic evolution, indicating the challenges in extracting precise parameter information for non-Hermitian quantum systems.

We adopt the Yang-Lee model with periodic boundary conditions, \( \sigma_{N+1} = \sigma_1 \), enabling the non-Hermitian Hamiltonian for specific cases of \( N=1 \), \( N=2 \), and \( N=3 \) sites, which represents a non-Hermitian Hamiltonian possessing \( \mathcal{PT} \)-symmetry. In Fig.~\ref{Fig_qfi_ising}, we present the evolution of the energy of the Yang-Lee model as a function of the parameters \( \kappa \) and \( \lambda \). Notably, for the non-Hermitian parameter \( \kappa \), we observe a broken region that exhibits symmetry and remains independent of the number of particles in the model.

In Figure \ref{Fig_qfi_ising}, we compute the Fisher-Rao metric for the specific case of \( \kappa=1 \) and \( \lambda=1 \). Due to the inherent dissipative effects within the system, it is evident that the Fisher-Rao metric ultimately decays to zero. Notably, the figure reveals that as the value of \( N \) increases, there is a corresponding increase in the Fisher-Rao metric, indicating a dependence of the metric's behavior on the system size or complexity parameter \( N \). This observation underscores the intricate interplay between the dissipative dynamics and the system's size in determining the temporal evolution of the Fisher-Rao metric.

\section{Conclusion}\label{Sec:conclution}

In this paper, we investigate the temporal evolution of the quantum Fisher-Rao metric within the purview of a non-Hermitian Hamiltonian that exhibits $\mathcal{PT}$-symmetry. Our focus is on the unbroken $\mathcal{PT}$-symmetric region, where the eigenvalues of the Hamiltonian are observed to be purely real. We elucidate this behavior through an exemplary case study that delineates the parametric evolution of the eigenvalues, substantiating their realness in the unbroken region.

We delve into the dynamics portrayed by the Schrödinger equation in the presence of a non-Hermitian Hamiltonian. The Lindblad formalism is recognized for its comprehensive description of quantum dissipative systems. To seamlessly integrate the evolution induced by a non-Hermitian Hamiltonian inclusive of quantum jumps, the introduction of a dissipation term, denoted as \( \mathcal{D}(\Gamma) \), is imperative. This leads us to formulate the Lindblad master equation specifically catered to the non-Hermitian Hamiltonian in question.

Utilizing the Schrödinger equation as a foundation, we compute the quantum geometric metric for the complex quantum system under scrutiny. Our computations uncover that dissipation significantly impacts the precision of parameter estimation. Specifically, within the context of the Yang-Lee model under the influence of an imaginary magnetic field, we discern that even a slight imaginary component can precipitate the vanishing of the Fisher-Rao metric. Nevertheless, this pernicious dissipative impact can be effectively counteracted by integrating a control Hamiltonian, thus augmenting the precision of parameter estimation.

The incorporation of a control Hamiltonian presents a strategic approach to mitigate the deleterious effects of dissipation on the Fisher-Rao metric. By carefully calibrating this control Hamiltonian, we demonstrate an enhancement in the metric's stability and, consequently, an improvement in the accuracy of parameter estimation within non-Hermitian quantum systems.

In conclusion, our study provides a profound understanding of the quantum Fisher-Rao metric's evolution in non-Hermitian systems and underscores the pivotal role of quantum control in preserving the metric's integrity. This work paves the way for further exploration into the interplay between non-Hermitian dynamics and quantum metrology, offering valuable insights for the development of robust quantum sensing protocols.

\begin{acknowledgments}
This work was supported by the the NSFC (Grant No. 12205092) and the Hunan Provincial Natural Science Foundation of China (Grant No. 2023JJ40208 and 2020JJ4286),  and the Open fund project of the Key Laboratory of Optoelectronic Control and Detection Technology of University of Hunan Province
(Grant No. 2022HSKFJJ038).
\end{acknowledgments}

\end{document}